\documentclass[10pt,twocolumn,letterpaper]{article}

\usepackage{cvpr}              

\usepackage[pagebackref,breaklinks,colorlinks,allcolors=cvprblue]{hyperref}
\usepackage{amssymb}
\usepackage{amsmath}
\usepackage{natbib}
\usepackage{tabularx}
\usepackage{multirow}
\usepackage{booktabs}
\usepackage{hyperref}
\usepackage{xcolor}
\usepackage[acronym]{glossaries}
\usepackage{xcolor}
\usepackage{tcolorbox}
\usepackage{array} 
\usepackage{booktabs} 
\usepackage{array}    
\usepackage{makecell} 
\newacronym{dlr}{DLR}{German Aerospace Center}
\newacronym{gsoc}{GSOC}{German Space Operations Center}
\newacronym{cfd}{CFD}{Concurrent Design Facility}
\newacronym{llm}{LLM}{Large Language Model}
\newacronym{plm}{PLM}{Pre-train Language Model}
\newacronym{rag}{RAG}{Retrieval Augmented Generation}
\newacronym{dl}{DL}{Deep Learning}
\newacronym{mmteb}{MMTEB}{Massive Multilingual Text Embedding Benchmark}
\newacronym{cdf}{CDF}{Concurrent Design Facility}
\newacronym{mlp}{MLP}{Multi Layer Perception}
\newacronym{calm}{CALM}{Concept Alignment and Latent Manipulation}
\newacronym{cw}{CW}{Concept Whitening}
\newacronym{svd}{SVD}{Singular Value Decomposition}
\newacronym{profs}{ProFS}{Projection Filter for Subspaces}
\newacronym{uwr}{UWR}{Unsafe Win Rate}
\newacronym{ndcg}{NDCG}{Normalize Discounted Cumulative Gain}
\newacronym{esa}{ESA}{European Space Agency}
\newacronym{mrr}{MRR}{Mean Reciprocal Rank}
\newacronym{BAAI}{BAAI}{Beijing Academy of Artificial Intelligence}
\newacronym{SLM}{SLM}{Small Language Model}
\newacronym{AI}{AI}{Artificial intelligence}
\newacronym{RAG}{RAG}{Retrieval Augmented Generation}
\newacronym{NLP}{NLP}{Natural Language Processing}
\newacronym{GPT}{GPT}{Generative Pre-train Transformer}
\newacronym{BERT}{BERT}{Bidirectional Encoder Representations from Transformers}
\newacronym{MTEB}{MTEB}{Massive text Embedding Benchmark}
\newacronym{KNN}{KNN}{K-Nearest Neighbor}
\newacronym{KG}{KG}{Knowledge graph}
\newacronym{IADC}{IADC}{Inter-Agency Space Debris Coordination Committee}
\newacronym{ESA}{ESA}{European Space Agency}
\newacronym{RL}{RL}{Reinforcement learning}
\newacronym{COT}{CoT}{chain-of-thought}
\newacronym{RNN}{RNN}{Recurrent Neural Networks}
\newacronym{LSTM}{LSTM}{Long Short-Term Memory}
\newacronym{FNN}{FNN}{Feed-forward neural network}
\newacronym{MLP}{MLP}{Multilayer perceptron}
\newacronym{MLM}{MLM}{Masked Language Model}
\newacronym{NSP}{NSP}{Next Sentence Prediction}
\newacronym{IR}{IR}{Information Retrieval}
\newacronym{TREC}{TREC}{Text Retrieval Conference}
\newacronym{DL}{DL}{Deep Learning}
\newacronym{Glove}{Glove}{Global Vectors for Word Representation}
\newacronym{DSSM}{DSSM}{Deep Structured Semantic Mode}
\newacronym{DAN}{DAN}{Deep Averaging Networks}
\newacronym{CNN}{CNN}{convolutional Neural Networks}
\newacronym{SPLADE}{SPLADE}{Sparse Lexical and Expansion model}
\newacronym{ColBERT}{ColBERT}{Contextualized Late Interaction over BERT}
\newacronym{MoE}{MoE}{Mixture of Experts}
\newacronym{SSM}{SSM}{state space model}
\newacronym{NDCG}{NDCG}{Normalize Discounted Cumulative Gain}
\newacronym{MAP}{MAP}{Mean Average Precision}
\newacronym{MRR}{MRR}{Mean Reciprocal Rank}
\newacronym{MSE}{MSE}{Mean Square Error}
\newacronym{BM25}{BM25}{Best Match 25}
\newacronym{TF/IDF}{TF/IDF}{Term-Frequency/Inverse-Document-Frequency}
\newacronym{qa}{Q\&A}{Question and Answers}

\definecolor{cvprblue}{rgb}{0.21,0.49,0.74}
\definecolor{promptcolor}{rgb}{0.95, 0.95, 0.95}
\newtcolorbox{promptboxcollumn}{
    colback=promptcolor,
    colframe=black,
    arc=5mm,
    boxrule=0.5pt,
    width=\columnwidth,
}

\def\paperID{23} 
\def\confName{CVPR}
\def\confYear{2026}

\title{A Systematic Evaluation of Retrieval-Augmented Generation and Language Models for Space Operations}

\author{
  Ruben Belo$^{1}$\thanks{Corresponding author: rc.belo@campus.fct.unl.pt}
  \qquad
  Marta Guimarães$^{2,1,3}$
  \qquad
  Cláudia Soares$^{1}$
  \\[1em]
  $^{1}$NOVA LINCS, Costa da Caparica, Almada, Portugal\\
  $^{2}$Neuraspace, Coimbra, Portugal\\
  $^{3}$Technical University of Munich, Germany\\[0.5em]
  {\tt\small rc.belo@campus.fct.unl.pt} \quad
  {\tt\small marta.guimaraes@neuraspace.com} \quad
  {\tt\small claudia.soares@fct.unl.pt}
}

\begin{document}
\maketitle
\begin{abstract}
The rapid expansion of space activities has led to an unprecedented accumulation of technical documentation, operational guidelines, and scientific literature, creating challenges for timely decision-making in space operations. Effective management in space operations requires tools capable of efficiently processing vast and heterogeneous information sources.

This paper systematically evaluates the performance of \gls{rag} pipelines, combining \glspl{llm} with information retrieval techniques for extracting and synthesizing actionable knowledge from domain-specific documents. We compare various retrieval strategies, embedding models, and \gls{llm} answers to assess their impact on information accuracy, relevance, and reliability.

Our results demonstrate that \gls{rag} pipelines can significantly enhance knowledge access, reduce uncertainty, and support decision-making in complex space operations.
\end{abstract}    
\section{Introduction}
\label{sec:intro}

The growing complexity and cadence of modern space missions have led to a substantial increase in the amount of documentation that engineers and operators must handle. This material spans a wide range of sources, including engineering analyses, operational procedures, regulatory documentation, and scientific literature. Mission teams are therefore expected to make timely, high-impact decisions based on information that is both highly specialized and continuously evolving. Despite this, many operational workflows continue to depend on manual document inspection, legacy toolchains, and human expertise, resulting in processes that are inefficient, difficult to scale, and costly~\cite{llm_ai_agent}. As a consequence, information handling in mission operations remains time-intensive, error-prone, and cognitively demanding for personnel~\cite{llm_ai_agent, llm_eval_space_opp}.

An example of such documentation is \gls{cdf} reports, which typically span 200–300 pages and encompass a wide range of mission-related information. These documents commonly include detailed discussions of mission architecture, system integration, payload design, ground segment infrastructure, operational concepts, and risk assessments~\cite{spaceQa_2022}. Another use case was identified at the \gls{gsoc} of the \gls{dlr}, where operators must continuously analyze large volumes of technical data and react promptly to unanticipated events~\cite{llm_eval_space_opp}. 

In this context, the integration of \gls{llm}-based tools offers the potential to reduce cognitive load by providing rapid, context-aware information support for engineers during both training activities and live mission operations~\cite{llm_eval_space_opp}.

The examples mentioned above underscore the increasing need for intelligent systems that can navigate domain-specific documentation and deliver accurate, context-aware information in real time. As space operations become more data-intensive, rapid retrieval and synthesis of relevant knowledge is now a critical requirement for mission success, operational safety, and long-term sustainability ~\cite{spaceQa_2022, llm_eval_space_opp, llm_ai_agent}. Using \gls{NLP}, \glspl{llm}, and artificial agents can enhance communication and decision-making for space operators, prevent potential collisions, and improve satellite trajectories~\cite{AI_space_traffic_management}. Illustrating this potential, \gls{llm}-based agents have been explored for operational satellite tasks such as anomaly detection and real-time decision support~\cite{satelite_ai_agents}. Related work has also investigated the use of \gls{llm} and autonomous AI agents for spacecraft guidance and control, highlighting the expanding role of these systems across diverse space operations~\cite{carrasco2025_llm_spacecraft}.

\glspl{llm}, when combined with \gls{rag}, offer a promising solution to the challenge of efficiently accessing, interpreting, and synthesizing large volumes of domain-specific and mission-critical documentation~\cite{llm_eval_space_opp, AI_space_traffic_management, belo_rag_space_2025}. The effectiveness of \gls{rag} has also been demonstrated in other technical scientific domains, where grounding \gls{llm} outputs in authoritative sources improves factual accuracy and reliability and reduces hallucinations of the generated answers~\cite{quantum_rag_qa}.

\begin{figure*}[t]
    \centering
        \includegraphics[width=0.8\linewidth]{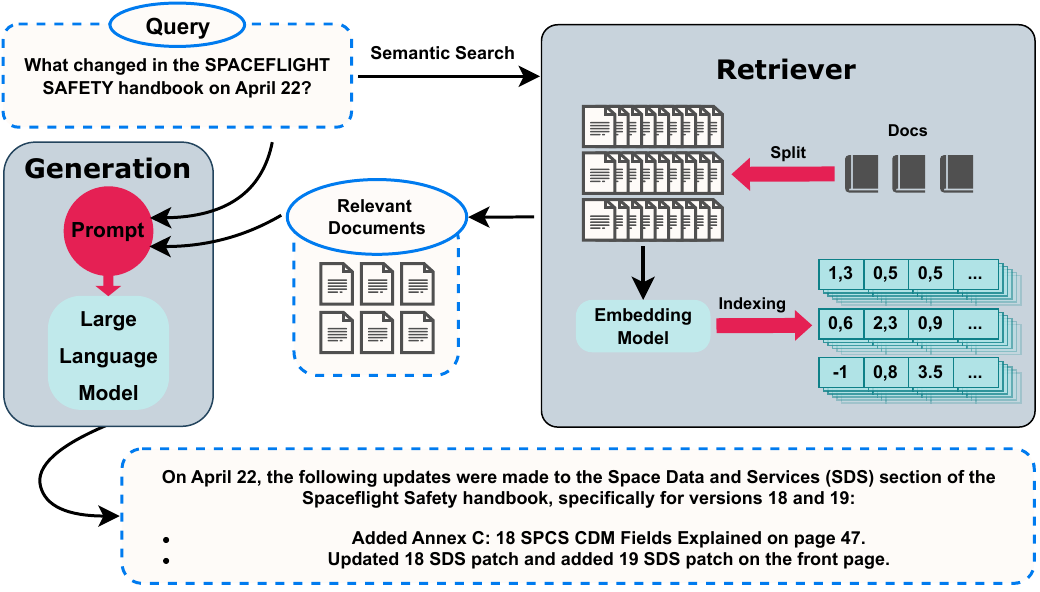}
    \caption[RAG example]{Overview of the information flow in a retrieval-augmented generation (RAG) pipeline for space-domain applications, illustrating document preprocessing and indexing, semantic retrieval of relevant passages, and their integration into a large language model for answer generation.}
    \label{fig:rag_example}
\end{figure*}

Finally, much of the data used in mission operations is highly sensitive and unsuitable for processing via cloud-based services due to security and confidentiality constraints~\cite{llm_eval_space_opp, satelite_ai_agents}. Accordingly, this work emphasizes open-source models and frameworks that can be deployed within closed and secure environments, providing full control over data privacy and operational execution both on the ground and in space.

\subsection{Related Work}
\label{sec:related_work}

The application of \glspl{llm} in space systems has attracted increasing attention as agencies face growing volumes of mission documentation. Early work adapted general-purpose models such as BERT~\cite{bert} to domain-specific contexts~\cite{spaceTransformers} addressing the limitations of small, specialized corpora by further pretraining three models on curated space datasets relevant to space operations.

Open-domain question answering tailored to space missions has also been explored. SpaceQA~\cite{spaceQa_2022} combined dense retrieval with neural reading to extract information from mission design documentation, highlighting the importance of domain-specific fine-tuning. Complementary efforts investigated \gls{qa} over structured knowledge bases, translating natural language queries into formal database programs~\cite{DISCOSQA_KB_QA_2023}.

Beyond textual \gls{qa}, multimodal approaches have been investigated, such as vision-language models fine-tuned on Martian terrain, demonstrating the potential of foundation models for scalable and autonomous space operations~\cite{AI4Mars_FM_2024}. 

In operational settings, \glspl{llm} have been evaluated for rapid information retrieval and decision support at \gls{gsoc}~\cite{llm_eval_space_opp}. They have also been studied as expert agents for mission management, performing tasks such as data analysis, summarization, and planning under offline deployment constraints to ensure confidentiality~\cite{llm_ai_agent}.

Across these domains, a recurring challenge is the integration of external knowledge into \glspl{llm}. \gls{rag} has emerged as a widely adopted approach to improve factual accuracy and ensure adequate domain coverage, supporting both \gls{qa} and operational decision-making~\cite{rag_survey, llm_eval_space_opp, llm_ai_agent}. In this context, ~\citet{belo_rag_space_2025} present an initial evaluation of \gls{rag}-based approaches for space operations, comparing two retrieval models and providing a brief qualitative analysis of the answers generated by an \gls{llm} using synthetically generated questions derived from space debris mitigation documents.

\subsection{Our Work}
\label{sec:our_work}
In this paper, we extend the evaluation of key components in \gls{rag} pipelines applied to space-specific text, building on previous work by~\citet{belo_rag_space_2025}. Consistent with recent studies~\cite{llm_eval_space_opp, belo_rag_space_2025}, the retrieval module (explained in Sec. \ref{sec:rag}) is of utmost importance, as the quality of the retrieved passages directly influences the quality of the generated answers. Thus, this paper analyzes retrieval models, providing insights into critical design decisions while ensuring strong retrieval effectiveness. Furthermore, we evaluate complete \gls{rag} pipelines not only to assess the factual accuracy of the generated answers, complementing prior work~\cite{belo_rag_space_2025}, but also to analyze answer quality. Given the heterogeneity of subdomains in this field, our study focuses on space debris mitigation documentation and the \gls{esa} mission \gls{qa} dataset as representative subdomains of space operations.

    

Our main contributions are:

\begin{enumerate}

    \item We extend the previous dataset in \cite{belo_rag_space_2025} by adding negative passage samples per query, thereby creating a context relevance dataset (Sec.~\ref{sec:pos_neg_mining}).

    \item We evaluate three reranking models for context relevance classification, demonstrating their effectiveness in distinguishing relevant from irrelevant passages and validating their use for retrieval evaluation (Sec.~\ref{sec:part_1_rerank_eval}).  

    \item We benchmark eight state-of-the-art embedding models and BM25~\cite{bm25} as retrievers within our evaluation framework using four metrics (Secs.~\ref{sec:part_1_eval_embed_models}). We further analyze relevance scores assigned by \gls{llm} over retrieved and reranked passages (Secs.~\ref{sec:embed_vs_rerank}). Both scenarios are evaluated using different chunking strategies (512 vs.\ 2.000 tokens) to assess their impact.



    \item We assess the factual accuracy of generated answers using a curated dataset from the \gls{esa} and further evaluate answer quality (Sec.~\ref{sec:part_1_answer_accuracy_spaceQA}).  

\end{enumerate}

\section{How to Take Advantage of \gls{rag} to Improve \glspl{llm} Answers?}
\label{sec:rag}

\glspl{llm} have demonstrated strong performance across a wide range of language tasks~\cite{llm_industry_lens,llm_research,llm_widespread_writing, quantum_rag_qa}, which justifies their increasing use across diverse applications. However, they continue to struggle with domain-specific or knowledge-intensive tasks~\cite{llm_strugle}. \gls{rag} addresses this limitation by incorporating external knowledge relevant to the query as contextual grounding, thereby improving the reliability of \gls{llm}-generated answers~\cite{rag_survey}.


The baseline \gls{rag} pipeline consists of two main components: the \textbf{Retriever} and the \textbf{Generator}~\cite{rag_survey}. 
The retriever identifies information that is likely to be useful for answering a given query. 
The generator uses this information to produce a coherent and informative response. 
The architecture and information flow of a typical \gls{rag} pipeline can be seen in Fig.~\ref{fig:rag_example}.

To start, documents (e.g., guidelines, articles, papers, books, or reports) are first preprocessed by splitting them into smaller, semantically meaningful passages, called \emph{chunks}. 
These chunks are then embedded and indexed. 
Here, embedding refers to transforming each text segment into a dense numerical vector that captures its semantic meaning, typically using a pretrained embedding model.

When a user asks a question, the retriever embeds the query in the same vector space, allowing it to select the most relevant passages based on cosine similarity,
$
\mathrm{sim}(\mathbf{q}, \mathbf{p}) = \frac{\mathbf{q} \cdot \mathbf{p}}{\|\mathbf{q}\| \, \|\mathbf{p}\|},
$
where $\mathbf{q}$ and $\mathbf{p}$ denote the query and passage embeddings, respectively. $\mathrm{sim}(\mathbf{q}, \mathbf{p})$ yields a value in the range $[-1, 1]$, where $-1$ denotes the most dissimilar pair and $1$ the most similar.

These retrieved passages are then provided as contextual input to an \gls{llm}, which combines them with the user query to generate a final answer. By leveraging this additional context, \gls{rag} systems combine the language understanding capabilities of \glspl{llm} with external, up-to-date information, without requiring model retraining. This makes \gls{rag} a valuable approach for improving both the accuracy and credibility of generated answers when compared to generation without retrieval~\cite{rag_survey} (Sec.\ref{sec:part_1_answer_accuracy_spaceQA}).

Beyond the core components of a \gls{rag} pipeline, additional modules may be incorporated (e.g., query rewriting, summarization, hybrid retrieval mechanisms, or prompt construction techniques)~\cite{rag_survey}. One such component is the \textbf{Reranker}, a deep learning language model that reorders retrieved passages by prioritizing the most relevant passage chunks earlier in the prompt and filtering out irrelevant ones, improving retrieval quality. However, this improvement comes at a higher computational cost~\cite{rag_survey}, making rerankers unsuitable as direct replacements for retrievers in large-scale systems. In this work, we instead leverage rerankers as a tool to help us understand retriever performance.

\section{Why and How We Evaluate the Retriever}
\label{sec:why_eval_retriever}

We focus on the retriever, as it largely determines overall system performance due to its direct impact on downstream generation tasks~\cite{better_rag}: without relevant and accurate context, faithful and robust responses are impossible.

In order to assess the retriever systematically without an in-domain dataset for this task, we rely on strong open-source context relevance classifiers (\textbf{Rerankers}), as a proxy relevance signal for ground truth. By doing so, we can approximate how well a retriever captures the passages that matter most for answering a query.
\begin{figure*}[t]
    \centering
        \includegraphics[width=0.8\linewidth]{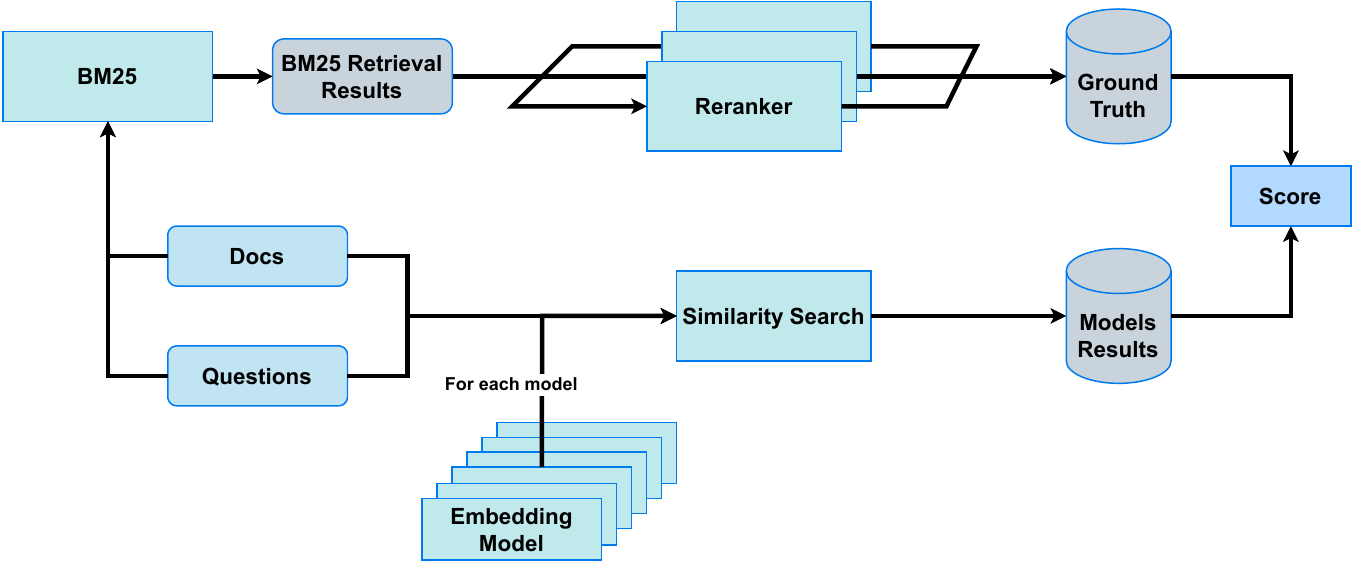}
    \caption[Embedding Models Evaluation Schema]{Embedding Models Evaluation Framework, where for each question we compute multiple passages rankings using the selected embedding models. These rankings are then compared with the reranker’s ranking, obtained after an initial retrieval step with BM25. In addition, we evaluate the retrieval capabilities of BM25 against the reranking.}
    \label{fig:eval_retriever_rerank}
\end{figure*}

\section{Context Relevance Dataset and Negative Mining}
\label{sec:pos_neg_mining}

To assess the suitability of open-source rerankers for our domain, we employ a dataset consisting of query-relevant passage pairs, with each query paired with two irrelevant passages. We build on previous work which created a dataset containing query-passage pairs from literature related to space debris mitigation ~\cite{belo_rag_space_2025}. We complement this by generating negative samples, passages not relevant to the query. To generate negative samples for our tasks, similar to the approach in \cite{ARES}, we employ two strategies:
\begin{itemize}

    \item \textbf{Random Negative Sampling:} We randomly sample in-domain passages that are unrelated to the source document of the synthetic query. Passage irrelevance is assessed using \texttt{BGE-M3}~\cite{bge-m3}, from which we sample the least $k$ relevant passages according to the model's scoring.

    \item \textbf{In-Document Negative Sampling:} This strategy follows a similar procedure to random negative sampling, but instead, we sample in-domain passages from the same document as the original passage. These passages are typically more topically similar, making them harder negatives for the model to distinguish. Although the sampled negatives can still be relatively easy, since passages that are semantically subtle yet lexically similar are not explicitly targeted.

\end{itemize}

\subsection{Why it is not recommended to use this dataset directly for Retrieval Evaluation}
There are several reasons why this dataset should not be used directly to evaluate retrieval methods. In synthetic datasets, the passage most relevant to a given query is not always the same as the passage from which the query was originally generated~\cite{gecko}. This misalignment makes it difficult to rely on origin passages as absolute ground truth.

Second, the dataset limits the types of evaluation metrics that can be applied. For example, metrics such as \gls{ndcg}, which depend on ranking multiple relevant passages and their scores, cannot be used reliably. This issue also affects other retrieval metrics, since retrievers may correctly prioritize different but still relevant passages. 

To address this, we argue that the dataset is better suited for evaluating rerankers. Rerankers provide relevance scores over candidate passage sets, enabling the construction of ranked relevance judgments that can be used as a more reliable ground truth for retriever evaluation. A similar strategy was applied in \citet{gecko}, where an \gls{llm} was used to analyze the top-$k$ retrieved passages and select the most relevant passage for a synthetic query to further train an embedding model. In our case, we instead use rerankers, as they provide intrinsic relevance scores for each query-passage pair that can be directly exploited for this purpose, while typically being more computationally efficient, with models often having one to two orders of magnitude fewer parameters.
\section{Evaluating Retrieval Models}

\subsection{Validating Rerankers}
\label{sec:part_1_rerank_eval}

\begin{table}[h]
\centering
\renewcommand{\arraystretch}{1} 
\caption{F1-Score and Accuracy comparison of BGE-M3, GTE-reranker-base, and Jina-reranker-v2 on the Context Relevance dataset.}
\resizebox{\columnwidth}{!}{%
\begin{tabular}{l*{6}{c}}
\toprule
\multirow{2}{*}{\textbf{Subset}} 
& \multicolumn{2}{c}{\textbf{BGE-M3}} 
& \multicolumn{2}{c}{\textbf{GTE}} 
& \multicolumn{2}{c}{\textbf{Jina}} \\
\cmidrule(lr){2-3} \cmidrule(lr){4-5} \cmidrule(lr){6-7}
& {F1-Score} & {Accuracy} 
& {F1-Score} & {Accuracy} 
& {F1-Score} & {Accuracy} \\
\midrule
Golden-Offset & .9395 & .9463 & .9568 & .9618 & .9543 & .9602 \\
Golden-Aligned & .9310 & .9390 & .9480 & .9540 & .9300 & .9399 \\
\bottomrule
\end{tabular}%
}
\label{tab:reranker_f1_acc}
\end{table}

We selected three rerankers: 
\href{https://huggingface.co/BAAI/bge-reranker-v2-m3}{BGE-M3 reranker}, 
\href{https://huggingface.co/Alibaba-NLP/gte-multilingual-reranker-base}{GTE-reranker-base}, and 
\href{https://huggingface.co/jinaai/jina-reranker-v2-base-multilingual}{Jina-reranker-v2}. 
Using multiple rerankers rather than a single model helps mitigate biases\footnote{Here, bias refers to systematic preferences introduced by model architecture, training data, or objective functions that may favor specific embedding models or textual patterns, potentially skewing relevance judgments.}. Additionally, because some rerankers are developed by the same organizations as certain embedding models, their outputs may inadvertently favor those models. To mitigate this potential bias, we use multiple rerankers in an ensemble, providing a more neutral assessment of retrieval performance.


Table~\ref{tab:reranker_f1_acc} reports F1-Score and Accuracy for each reranker. 
\textbf{Accuracy} measures the proportion of correctly classified contexts (relevant or irrelevant) and can be calculated as $\text{Accuracy} = \frac{\# \text{correct predictions}}{\# \text{total predictions}}
$. 
\textbf{F1-Score} is the harmonic mean of precision and recall, given by $\text{F1-Score} = 2 \cdot \frac{\text{Precision} \cdot \text{Recall}}{\text{Precision + Recall}}$, where $\text{Precision} = \frac{\text{TP}}{\text{TP + FP}}$ and $\text{Recall} = \frac{\text{TP}}{\text{TP + FN}}$, with TP, FP, and FN denoting true positives, false positives, and false negatives.

As shown, all rerankers demonstrate strong classification performance, indicating reliable distinction between relevant and irrelevant contexts, supporting their use to evaluate retrieval outputs in downstream tasks.

\subsection{Evaluation of Embedding Models}
\label{sec:part_1_eval_embed_models}

To start our evaluation, we selected eight state-of-the-art embedding models from the \gls{mmteb} leaderboard~\cite{MMTEB_2025}. These embedding models can be commonly used as retrievers, and the leaderboard benchmarks them across a diverse range of tasks, serving as a standard reference in the field for evaluating new state-of-the-art models. The selection criteria, along with an overview of the embedding models used, are provided in the supplementary material. In addition, we evaluated BM25~\cite{bm25} as a widely adopted retrieval method and to build our ground truth. 

For our evaluation, we retrieved the top-$50$ passages from the entire corpus for each query using each method and model. As a proxy for ground truth, we applied a two-step procedure: first, retrieving the top-$100$ passages with BM25~\cite{bm25}, and then classifying each passage as relevant and reranking using the rerankers. This reranked list served as our ground truth. A visualization of the procedure can be seen in Fig.~\ref{fig:eval_retriever_rerank}.

\begin{figure}[h]
    \centering
        \includegraphics[width=1\linewidth]{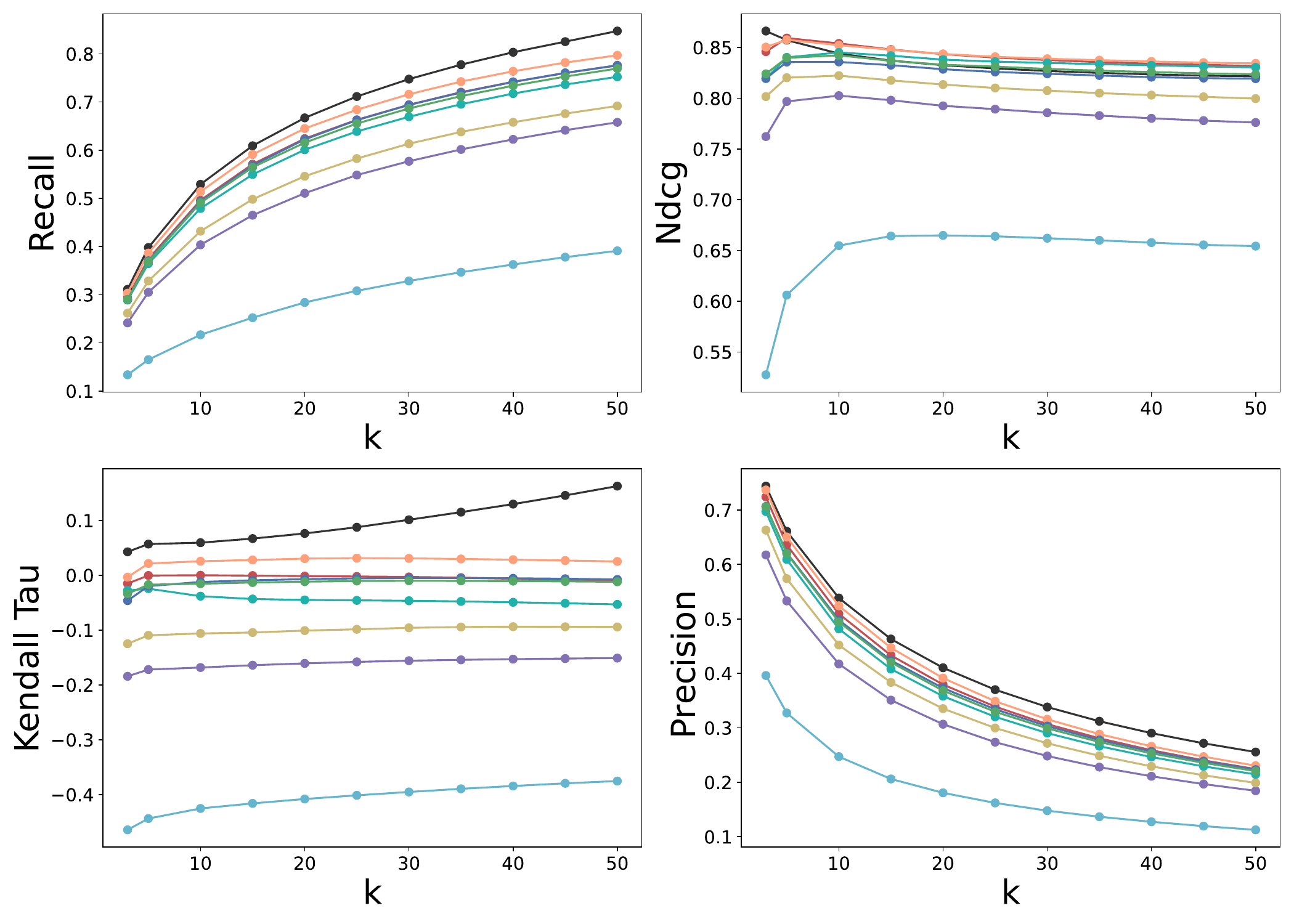}
        \includegraphics[width=1\linewidth,  trim=50 95 50 95, clip]{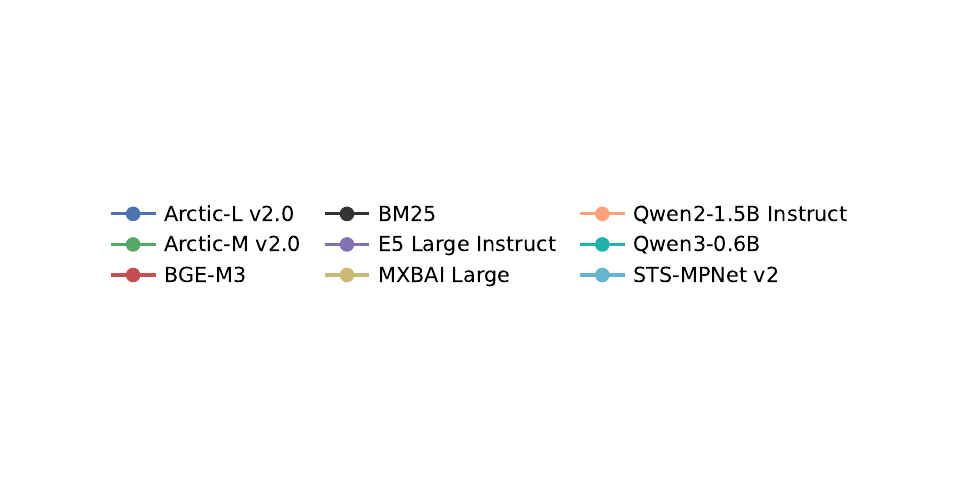}
    \caption[Average performance of embedding models, 2.000 token chunks]{Average embedding model performance across retrieval metrics with 2.000-token passage chunks, averaged across three rerankers. BM25~\cite{bm25} and Qwen2-1.5B Instruct ~\cite{qwen_embed} shows the most consistent performance across all metrics and $k$ values. STS-MPNet v2 and E5 Large perform the worst, with the lowest scores across all metrics. The remaining models achieve comparable results, clustering closely across evaluation criteria.}

    \label{fig:Embed_model_eval_2k}
\end{figure}

We computed four distinct metrics: Recall and Precision, calculated using the relevant passages, and \gls{ndcg} and Kendall Tau correlation, computed using the reranked list. Each metric was evaluated across multiple top-$k$ thresholds. To assess the impact of document chunking on retrieval performance, we conducted the evaluation with two chunking strategies: one with a maximum chunk size of 2.000 tokens\footnote{Tokens refer to the subword units produced by the tokenizer during the language model preprocessing step, rather than words or characters.}. and another with 512 tokens. Results for the 2.000 token setting are shown in Fig.~\ref{fig:Embed_model_eval_2k}, while the results for the 512 token setting are provided in the supplementary material Fig.~5. The two figures show relatively similar behavior between models across the different metrics.

\begin{table*}[th]
\centering
\caption{Relative frequency (\%) of relevance scores assigned by the Retriever and Reranker under 2.000-token and 512-token settings. Scores are on a 0-3 scale: 0 = completely irrelevant, 1 = slightly relevant, 2 = moderately relevant, and 3 = highly relevant. Results are reported for different Top-K retrieval ranges (3, 5, 7, 10).}

\label{tab:relevance_labels_by_llm_grouped}
\begin{tabular}{l l c c c c | c c c c}
\toprule
\multirow{2}{*}{\textbf{Top-K}} & \multirow{2}{*}{\textbf{Method}} 
& \multicolumn{4}{c|}{\textbf{2.000 tokens}} & \multicolumn{4}{c}{\textbf{512 tokens}} \\
\cmidrule(lr){3-6} \cmidrule(lr){7-10}
 &  & \textbf{0 $\downarrow$} & \textbf{1 $\downarrow$} & \textbf{2 $\uparrow$} & \textbf{3 $\uparrow$} 
 & \textbf{0 $\downarrow$} & \textbf{1 $\downarrow$} & \textbf{2 $\uparrow$} & \textbf{3 $\uparrow$} \\
\midrule
\multirow{2}{*}{Top-3} & Retriever & 2.12 & 3.44 & \textbf{54.78} & 39.66 & 1.48 & 3.92 & \textbf{52.06} & 42.54 \\
                        & Reranked  & \textbf{1.29} & \textbf{1.78} & 52.18 & \textbf{44.76} & \textbf{0.73} & \textbf{1.78} & 49.12 & \textbf{48.37} \\
\multirow{2}{*}{Top-5} & Retriever & 2.98 & 4.82 & \textbf{59.96} & 32.24 & 2.23 & 5.54 & \textbf{57.64} & 34.60 \\
                        & Reranked  & \textbf{2.13} & \textbf{2.76} & 58.55 & \textbf{36.56} & \textbf{1.17} & \textbf{2.71} & 56.05 & \textbf{40.07} \\
\multirow{2}{*}{Top-7} & Retriever & 3.78 & 5.82 & \textbf{62.52} & 27.88 & 2.93 & 6.70 & \textbf{60.39} & 29.99 \\
                        & Reranked  & \textbf{2.65} & \textbf{3.70} & 62.04 & \textbf{31.61} & \textbf{1.57} & \textbf{3.62} & 60.06 & \textbf{34.76} \\
\multirow{2}{*}{Top-10}& Retriever & 4.80 & 7.05 & 64.38 & 23.76 & 3.71 & 8.31 & 62.45 & 25.53 \\
                        & Reranked  & \textbf{3.51} & \textbf{4.80} & \textbf{64.87} & \textbf{26.82} & \textbf{2.13} & \textbf{4.78} & \textbf{63.46} & \textbf{29.63} \\
\bottomrule
\end{tabular}
\end{table*}

There are multiple ways to interpret these results. For instance, when designing a pipeline that includes a reranker, priority should be given to retrieval models with high recall and low latency (e.g., BM25), as the reranker is responsible for most of the relevance refinement. In such settings, maximizing the number of relevant passages passed to the reranker is critical, while efficiency remains important, as rerankers are significantly more computationally expensive than retrievers.

In contrast, when a pipeline relies solely on a retriever, greater emphasis should be placed on \gls{ndcg}, which accounts not only for passage relevance but also for the position at which relevant passages appear within the top-$k$ results, (e.g., BGE-M3 or Qwen 2).

\subsection{What is the Impact of Reranking? Chunk Size Analysis}
\label{sec:embed_vs_rerank}

Previous work evaluated the relevance of each passage for each query but
only with passages up to 2.000 chunk~\cite{belo_rag_space_2025}. They used Llama 3.3 70B~\cite{llama3} as the evaluator, since the use of \glspl{llm} for evaluation is already established in frameworks such as RAGAS~\cite{Ragas} and is increasingly adopted in advanced retrieval
setups~\cite{rag_survey}, where the model acts as a judge. Recent methods further demonstrate that \glspl{llm} can reliably determine which documents to retain or discard,
as well as when to actively retrieve additional information in an agentic manner~\cite{rag_survey}. We follow the same steps but also evaluate passages up to 512 tokens to see if reranking meaningfully improves the quality of retrieved content and the preference for the passage length.

We employ the same custom prompt (supplementary material Fig.~6) asking the model to score each question-passage pair on a scale from 0 to 3, with 0 meaning completely irrelevant and 3 meaning highly relevant. These multi-level relevance labels have been shown to improve the performance of \gls{llm}-based classifiers~\cite{scoresLabels} and we also evaluate four different top-k ranges: 3, 5, 7, and 10.

As shown in Table~\ref{tab:relevance_labels_by_llm_grouped}, several clear patterns emerge when comparing retriever and reranker outputs across both 2.000-token and 512-token settings.

First, reranking consistently reduces the proportion of irrelevant documents (scores 0 and 1) across all top-k values, regardless of chunk size. For instance, in the 2.000-token setting at Top-3, the reranker decreases the share of completely irrelevant documents from 2.12\% to 1.29\% and increases the proportion of completely relevant documents (score 3) from 39.66\% to 44.76\%. A similar effect is observed in the 512-token setting, where top-3 reranked outputs reduce 0-score documents from 1.48\% to 0.73\% and increase 3-score documents from 42.54\% to 48.37\%. This demonstrates that reranking effectively filters out less relevant passages while boosting highly relevant ones, independent of chunk size.

Second, for moderately relevant documents (score 2), the retriever generally achieves slightly higher proportions than the reranker, particularly at lower top-k values, indicating that the reranking step is most beneficial at lower top-k thresholds, where filtering out irrelevant content has a bigger impact on overall relevance. Third, comparing relevance scores across chunk sizes, the retriever with 2.000-token passages tends to have higher values for score 2, whereas the retriever with 512-token passages consistently shows higher values for score 3, this same trend also happens in the reranker. This aligns with the intuition that longer passages may contain more noise or irrelevant content, slightly diluting the relevance of top passages, and the \gls{llm} was a preference for this setup.

Overall the reranker with 512-token passages is the most consistently dominant setting. This suggests that, in practice, shorter, more focused passages combined with reranking provide the most effective retrieval setup, with the reranker leveraging the cleaner content of smaller chunks to improve the proportion of highly relevant documents.
\section{Evaluating LLM Answer Accuracy and Quality On SpaceQA}
\label{sec:part_1_answer_accuracy_spaceQA}

\subsection{Results}
\begin{table*}[ht]
\centering
\small
\caption{Evaluation of answer generation on the SpaceQA dataset with noisy context. Reported are average scores $\pm$ standard deviation, as well as accuracy with and without (w/o) context over the 60 examples.}
\label{tab:space_qa_llm_results}
\begin{tabular}{lccccc}
\toprule
 & \makecell{Answer \\ faithfulness} 
 & \makecell{Answer \\ relevance} 
 & \makecell{Noise \\ robustness} 
 & \makecell{Answer \\ accuracy} 
 & \makecell{Answer accuracy \\ w/o Context} \\
\midrule
\makecell{SpaceQA (Top-5)} 
& $3.92_{[3.30; 4.54]}$ & $4.02_{[3.31; 4.73]}$ & $4.52_{[3.62; 5.00]}$ & $56/60$ & $3/60$ \\
\bottomrule
\end{tabular}
\end{table*}

Previous work by \citet{belo_rag_space_2025} focused on evaluating the quality of answers generated by \glspl{llm} in the space domain. In contrast, to assess answer accuracy, we use the SpaceQA dataset created by \gls{esa}, which enables evaluation of a model’s ability to produce accurate answers. This dataset was previously used by \citet{spaceQa_2022} to assess the performance of the first open-ended question answering systems in the space domain.

This curated dataset consists of 60 question–passage–answer triplets produced by \gls{esa}. While its subdomain differs slightly from our primary focus on space debris mitigation, it includes detailed information on mission design, system configuration, payloads, service modules, ground segments, operations, and risk assessment, thereby serving as comprehensive references of \gls{esa} missions~\cite{spaceQa_2022}.

We examine whether Llama 3 8B~\cite{llama3} can generate accurate answers and how it performs under noisy conditions, where the context is corrupted but relevant information is still present.

\paragraph{Setup} To evaluate the answer-generation capabilities of our \gls{llm}, we provide the correct passage in the context for each question while also introducing noise by adding four random in-domain passages from the 2.000-token set. This setup maximizes Llama 3 8B~\cite{llama3} ratio of irrelevant to relevant tokens within the context. For evaluation, we measure \textbf{Answer Accuracy}, which quantifies the proportion of generated answers that match the available ground-truth answers in the dataset. In addition, we assess three metrics of answer quality that are commonly used in \gls{rag} evaluations: \textbf{Answer Faithfulness}, \textbf{Answer Relevance}, and \textbf{Noise Robustness}~\cite{rag_survey, belo_rag_space_2025}. Following prior work~\cite{belo_rag_space_2025,Ragas}, these metrics were evaluated using an ensemble of LLaMA 3.3 70B~\cite{llama3} and Qwen 2.5 72B~\cite{qwen2.5} as judges, with each metric scored on a 1–5 scale~\cite{scoresLabels}, where higher scores indicate better quality.

Answer faithfulness ($3.92$) and relevance ($4.02$) indicate that generated responses are generally accurate with respect to the provided evidence and align well with the questions. The relatively high robustness score ($4.52$) demonstrates that the model is able to filter out irrelevant information even when four noisy passages are included in the context. Importantly, the answer accuracy of $56/60$ shows that the model not only produces plausible text but also arrives at the correct ground-truth answers in most cases. In contrast, without RAG, accuracy drops drastically to $3/60$, highlighting the critical role of retrieved context in guiding the model toward correct answers.

Taken together, these results suggest that the \gls{llm} can leverage context effectively while remaining resilient to distractor passages. However, given the limited dataset of 60 examples, these findings should be interpreted as indicative rather than definitive.

A closer examination of the incorrect cases when context is provided reveals that most failures arise when the connection between the question and the supporting passage is weak or only implicit. In these instances, the passage often contains the relevant information but does not explicitly reference the entities or terms used in the question. Correctly answering, therefore, requires the model to perform bridging inferences or make assumptions that go beyond the literal text. The model, however, generally avoids such speculation and defaults to conservative responses, claiming that the answer is not present in the context. This cautious behavior reduces the risk of hallucination but limits the model’s ability to extract correct answers when indirect reasoning is required. To illustrate this limitation, we present the four incorrect examples in detail below.

\subsubsection{Analysis of Incorrect Answers from SpaceQA}

\begin{figure}[h]  
\centering
\begin{promptboxcollumn}
\textbf{Question:} 

\hspace{3mm}What is the size of the solar array on ATHENA? 

\textbf{Passage:} 

\hspace{3mm}This results in a total of 30m2 array area, 154 kg total mass. This is assumed to be implemented as 2 fixed deployable wings, each 15m2, 77 kg.

\textbf{Answer:}

\hspace{3mm}The context provided does not mention ATHENA or its solar array size. The text only discusses CubeSats, collision risk, and space data exchange standards.

\end{promptboxcollumn}
\caption[SpaceQA incorrect answer example 1]{Example \#1 of an incorrect answer generated by the model on the SpaceQA dataset. The model fails to extract the relevant information about ATHENA's solar array size from the provided passage, likely due to a weak semantic link between the question and passage content.}
\label{fig:space_qa_example_1}
\end{figure}

\paragraph{Example 1}

In Fig.~\ref{fig:space_qa_example_1}, the passage actually provides the solar array size (30 m² in total, split into two 15 m² wings), but the connection to ATHENA is implicit. Because ATHENA is not explicitly mentioned in the passage, the model treats the information as unrelated and refuses to attribute the array size to ATHENA. This shows the model’s reluctance to bridge entity references when they are not directly aligned with the question, favoring caution over inference.

\paragraph{Example 2}

In the supplementary material Fig.~7, the passage explicitly mentions a launch in 2028, which is the correct answer. However, because the passage frames the launch requirement indirectly and does not repeat the mission name “ATHENA,” the model overemphasizes contextual uncertainty. Instead of extracting the straightforward launch year, it digresses into a lengthy discussion of TRL readiness and cost frameworks before ultimately disclaiming that ATHENA is not mentioned. This indicates a failure to prioritize direct numerical answers when they are embedded in a broader technical discussion.

\paragraph{Example 3}

In the supplementary material Fig.~8, here the model focuses on the requesting and funding entities (ESA/SRE-FM and GSP) rather than the supporting contributors (JAXA and SRON experts). The correct supporting organizations are explicitly present in the passage, but the model fails to attend to the latter part of the text. This highlights a weakness in selective information retrieval: when multiple organizational roles are described in a passage, the model locks onto the first entities mentioned and disregards subsequent ones.

\paragraph{Example 4}

In the supplementary material Fig.~9, the passage does in fact state both a technical goal (demonstrating long-distance driving within a short timeframe) and a scientific goal (investigating multiple sites under similar conditions). The model fails to connect “this mission” to “MarsFAST,” perhaps due to a mismatch in naming: the mission name is not explicitly repeated in the passage. Consequently, the model defaults to claiming irrelevance. This shows that the model struggles with pronoun and co-reference resolution when mission names are not reiterated in the supporting text.

It is important to note that these limitations are not solely attributable to the \gls{llm} itself. Many of the observed incorrect answers arise from situations where critical information or entity mentions are not present in the necessary chunk, making it difficult for the model to answer correctly without making assumptions or adding additional context. Such issues could potentially be addressed with alternative chunking strategies or the inclusion of richer metadata that provides mission-specific information (e.g., mission name, mission type, or equipment used). Since the dataset was predefined, we had no control over the presence or distribution of this information. These observations highlight that retrieval design plays a critical role in overall system performance, and that careful consideration of these factors is necessary when evaluating and deploying \gls{llm}-based \gls{qa} systems for space-domain applications.

\section{Conclusion}

In this work, we systematically evaluated state-of-the-art models and their use in the principal components of a \gls{rag} pipeline with the specific focus in the space domain. More specifically, we evaluate 8 different embedding models as well as the BM25 algorithm as retrieval options. We also evaluate two different retrieval strategies with and without reranking and different passage lengths. With this we learn that some models are up to the task, the use of the reranker  is strongly beneficial to maximize context relevance as well as the preference for passages up to 512 tokens. We evaluate the generator with four complementary metrics, simulating from standard performance up to extreme system failure. We also complement a previously made dataset with passage-irrelevant passage triplets.

Our results show that current models are promising for controlled deployment in space operations, with some retrievers and rerankers demonstrating impressive performance, and with the \gls{llm} exhibiting both strong answer generation capabilities and robustness in scenarios with highly corrupted retrieval.

\section{Limitations}
The results presented here should be interpreted in context. The SpaceQA dataset contains only 60 Q\&A pairs, and while the results should be interpreted as indicative rather than definitive, this provides an initial benchmark but limits large-scale statistical conclusions; future work could use synthetic or augmented datasets, with caution, to expand evaluation. The in-document negative sampling strategy may include passages that are easy to distinguish, suggesting opportunities for constructing harder, more confusable negatives. Reranker outputs are used as ground-truth proxies, and while we ensemble multiple rerankers, the inclusion of more complex signals (e.g., LLM-based scoring or independent human judgments) could further validate results. Answer quality metrics rely on an ensemble of LLaMA 3.3 70B and Qwen 2.5 72B; while robust, alignment with human judgment remains an open challenge in RAG evaluation.

\section{Acknowledgments}
We thank the anonymous reviewers for their valuable input and for helping
strengthen the manuscript. This work was partially supported by NOVA LINCS (UID/04516) funded
by FCT IP, and the Neuraspace AI Fights Space Debris project (project code C626449889-00463050,
operation code 2022-C05i0101-02), co-funded by Recovery and Resilience Plan and NextGeneration
EU Funds, www.recuperarportugal.gov.pt.

{
    \small
    \bibliographystyle{ieeenat_fullname}
    \bibliography{main}
}
\clearpage

\appendix
\onecolumn
\newpage

\section{Rerank Evaluation extra Data}

\begin{table*}[h]
\centering
\caption{Classification performance comparison of BGE-M3, GTE-reranker-base, and Jina-reranker-v2 on the Context Relevance dataset.}
\resizebox{\textwidth}{!}{%
\begin{tabular}{ll*{9}{c}}
\toprule
\textbf{Setting} & \textbf{Label} 
& \multicolumn{3}{c}{\textbf{BGE-M3}} 
& \multicolumn{3}{c}{\textbf{GTE}} 
& \multicolumn{3}{c}{\textbf{Jina}} \\
\cmidrule(lr){3-5} \cmidrule(lr){6-8} \cmidrule(lr){9-11}
 & & {Precision} & {Recall} & {F1-Score} 
   & {Precision} & {Recall} & {F1-Score} 
   & {Precision} & {Recall} & {F1-Score} \\
\midrule
\multirow{3}{*}{Golden-Offset} 
 & Irrelevant Context & .9590 & .9604 & .9597 & .9669 & .9761 & .9715 & .9514 & .9909 & .9708 \\
 & Relevant Context   & .9206 & .9180 & .9193 & .9512 & .9332 & .9421 & .9802 & .8988 & .9378 \\
 \cmidrule(lr){3-5} \cmidrule(lr){6-8} \cmidrule(lr){9-11}
 & Macro Avg          & .9398 & .9392 & .9395 & .9591 & .9546 & .9568 & .9658 & .9449 & .9543 \\
 \cmidrule(lr){2-11}
 & Accuracy & \multicolumn{3}{c}{.9463} & \multicolumn{3}{c}{.9618} & \multicolumn{3}{c}{.9602} \\
\midrule
\multirow{3}{*}{Golden-Aligned} 
 & Irrelevant Context & .9498 & .9592 & .9544 & .9609 & .9704 & .9657 & .9272 & .9873 & .9563 \\
 & Relevant Context   & .9167 & .8986 & .9075 & .9397 & .9211 & .9303 & .9709 & .8451 & .9036 \\
 \cmidrule(lr){3-5} \cmidrule(lr){6-8} \cmidrule(lr){9-11}
 & Macro Avg          & .9332 & .9289 & .9310 & .9503 & .9458 & .9480 & .9491 & .9162 & .9300 \\
 \cmidrule(lr){2-11}
 & Accuracy & \multicolumn{3}{c}{.9390} & \multicolumn{3}{c}{.9540} & \multicolumn{3}{c}{.9399} \\
\bottomrule
\end{tabular}%
}
\label{tab:reranker_comparison}
\end{table*}

\section{Model Information and Selection Criteria}
\label{app:part_1_info_selec_criteria_embed_model}

\begin{table}[h]
\centering
\caption{Comparison of selected embedding models. Columns show: \textbf{Organization} (model developer), \textbf{Model Name}, \textbf{Parameters} (trainable weights), \textbf{Context Length} (maximum tokens per input), and \textbf{Embedding Dimension} (vector size). Performance metrics from the \gls{mmteb} leaderboard~\cite{MMTEB_2025} include: \textbf{Clustering} (grouping similar texts), \textbf{Reranking} (reordering retrieved results), \textbf{Retrieval} (finding relevant documents), and \textbf{STS} (Semantic Textual Similarity).}

\resizebox{\columnwidth}{!}{%
\begin{tabular}{lccccccccc}
\toprule
\textbf{Org.} & \textbf{Model} & \textbf{Params} & \textbf{Context} & \textbf{Dim} & \textbf{Clust.} & \textbf{Rerank} & \textbf{Retr.} & \textbf{STS} \\
\midrule
Gameselo & \href{https://huggingface.co/Gameselo/STS-multilingual-mpnet-base-v2}{STS‑MPNet v2} & 278M & 514 & 768 & 32.38 & 47.43 & 34.66 & 69.33 \\

intfloat & \href{https://huggingface.co/intfloat/multilingual-e5-large-instruct}{E5 Large Instruct} & 560M & 514 & 1024 & 50.75 & \textbf{76.81} & 62.61 & 57.12 \\

Qwen & \href{https://huggingface.co/Qwen/Qwen3-Embedding-0.6B}{Qwen3‑0.6B} & 595M & 32768 & 1024 & \textbf{52.33} & 60.26 & \textbf{64.65} & \textbf{74.94} \\

Alibaba-NLP & \href{https://huggingface.co/Alibaba-NLP/gte-Qwen2-1.5B-instruct}{Qwen2‑1.5B Instruct} & 1B & 32768 & 8960 & 52.05 & 61.61 & 60.78 & 70.36 \\

BAAI & \href{https://huggingface.co/BAAI/bge-m3}{BGE‑M3} & 568M & 8194 & 1024 & 40.88 & 61.98 & 54.60 & 72.99 \\

Snowflake & \href{https://huggingface.co/Snowflake/snowflake-arctic-embed-m-v2.0}{Arctic‑M v2.0} & 305M & 8192 & 768 & 42.24 & 60.45 & 54.83 & 65.91 \\

Snowflake & \href{https://huggingface.co/Snowflake/snowflake-arctic-embed-l-v2.0}{Arctic‑L v2.0} & 568M & 8192 & 1024 & 42.76 & 62.89 & 58.36 & 69.48 \\

mixedbread‑ai & \href{https://huggingface.co/mixedbread-ai/mxbai-embed-large-v1}{MXBAI Large} & 335M & 512 & 1024 & 42.49 & 44.51 & 40.30 & 61.20 \\

\bottomrule
\end{tabular}%
}
\label{tab:embedding_models}
\end{table}

Our selection criteria focused on open-source models with fewer than 2 billion parameters, ensuring diversity in model size, embedding dimensions, as well as differences in architecture and spanning different research groups and organizations. The selected embeddings can be seen in Table \ref{tab:embedding_models}.

\newpage

\section{Evaluating Retrieval Models Extra}
\label{app:part_1_embed_512_results}

\begin{figure}[htbp]
    \centering
    \includegraphics[width=0.9\linewidth, trim=10 10 10 10, clip]{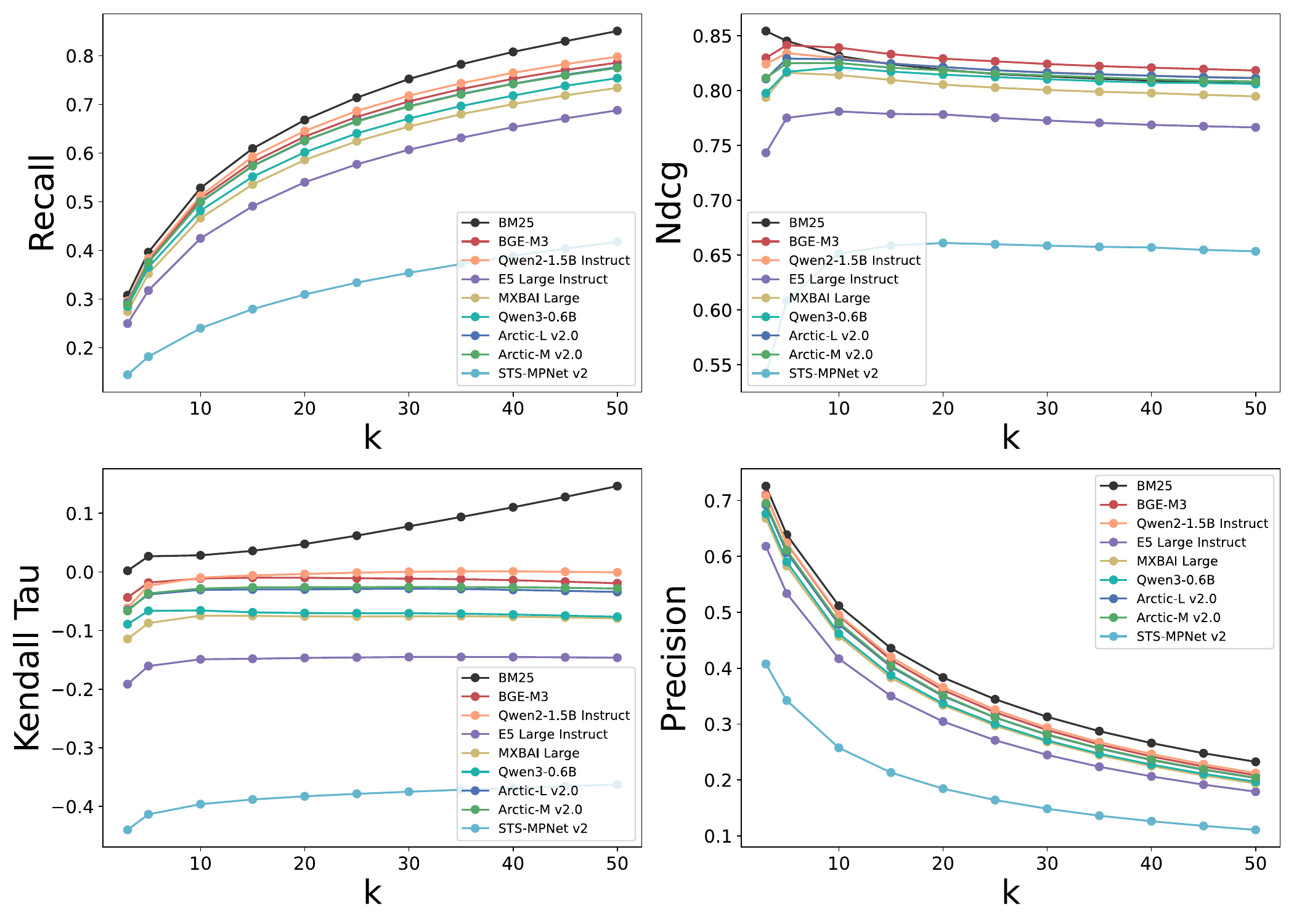}
    \includegraphics[width=0.8\linewidth, trim=50 95 50 95, clip]{figs/legend_only.pdf}

    \caption[Embedding model performance, 512-token chunks]{\textbf{Average Embedding model performance across retrieval metrics with 512-token document chunks:} BM25 and Qwen2-1.5B Instruct show the most consistent performance across all metrics and $k$ values. 
STS-MPNet v2 and E5 Large perform worst, with the lowest scores across all metrics. 
The remaining models achieve comparable results, clustering closely across evaluation criteria.}
    \label{fig:Embed_model_eval_512}
\end{figure}

\newpage

\section{Prompts used for evaluation}

\begin{figure}[h]  
\centering
\begin{promptboxcollumn}
System prompt:

\hspace{5mm}You are an AI assistant that judges the relevance of a document to a given question. Respond with a score from 0 to 3.

\hspace{5mm}QUESTION: {User question}

\hspace{5mm}DOCUMENT: {doc}

\hspace{5mm}Rate how relevant the document is to answering the question using the following scale:

\hspace{10mm}0 = Completely irrelevant

\hspace{10mm}1 = Slightly irrelevant

\hspace{10mm}2 = Slightly relevant

\hspace{10mm}3 = Completely relevant

\hspace{5mm}Respond with a single number between 0 and 3.

\end{promptboxcollumn}
\caption[Context Relevance Prompt]{\textbf{Context Relevance Prompt:} Used to test both the Retriever and the Reranker.}
\label{fig:doc_relevance_prompt}
\end{figure}

\section{Incorrect Answers in SpaceQA}
\label{app:wrong_spaceqa}

\begin{figure}[h]  
\centering
\begin{promptboxcollumn}
\textbf{Question:} 

\hspace{3mm}When will ATHENA be launched? 

\textbf{Passage:} 

\hspace{3mm}The main challenge for this project is to find technical solutions fitting to the cost frame work. Next to that all technologies with low TRL need to be advance such that they reach TRL 6 (RD[47]) before the start of the implementation phase. This could become difficult, although the launch is only required in 2028, because some parts of the X-Ray Field Unit are reported to be only at TRL 2 presently.

\textbf{Answer:}

\hspace{3mm}Based on the provided context, it seems that the main challenge is to find technical solutions that fit within a cost framework and advance technologies with low TRL (Technology Readiness Level) to TRL 6 before the implementation phase. This is important because some parts of the X-Ray Field Unit are currently at TRL 2 and need to be advanced before the launch in 2028.

To address this challenge, it would be helpful to identify the specific technologies that need to be advanced and prioritize them based on their potential impact on the project's success. Additionally, it may be necessary to allocate additional resources or funding to support the development of these technologies.

It's also important to note that the project has a specific timeline and budget constraints, so it's crucial to manage the project's scope, schedule, and budget effectively to ensure its successful completion.

Regarding the question about ATHENA's launch, it's not mentioned in the provided context. ATHENA is not a project mentioned in the document, so it's not possible to provide an answer to this question based on the provided information.

\end{promptboxcollumn}
\caption[SpaceQA incorrect answer example 2]{Example \#2 of an incorrect answer generated by the model on the SpaceQA dataset. The passage explicitly mentions a launch in 2028, but the model fails to extract this direct numerical information and instead focuses on contextual uncertainties.}

\label{fig:space_qa_example_2}
\end{figure}
\begin{figure}[h]  
\centering

\begin{promptboxcollumn}
\textbf{Question:} 

\hspace{3mm}Who supported ESA in the NG-CryoIRTel study? 

\textbf{Passage:} 

\hspace{3mm}The study was requested by ESA Science Directorate SRE-FM and funded by the General Studies Programme GSP. The study was carried out in 8 sessions, starting with a kick off on the 13th November 2014 and ending with an Internal Final Presentation on the 18th December 2014 by an interdisciplinary team of specialists from ESTEC and ESOC and supported by SPICA/SAFARI experts from JAXA and SRON.

\textbf{Answer:}

\hspace{3mm}The study was requested by ESA Science Directorate SRE-FM and funded by the General Studies Programme GSP.

\end{promptboxcollumn}
\caption[SpaceQA incorrect answer example 3]{Example \#3 of an incorrect answer generated by the model on the SpaceQA dataset. The model fails to identify the entities supporting ESA in the NG-CryoIRTel study and instead reports who requested and funded the study, demonstrating confusion between different organizational roles described in the passage.}
\label{fig:space_qa_example_3}
\end{figure}

\begin{figure}[h]  
\centering
\begin{promptboxcollumn}
\textbf{Question:} 

\hspace{3mm}What is the goal of the MarsFAST mission?

\textbf{Passage:} 

\hspace{3mm}The technical demonstration goal of this mission, to cover a long distance drive within a few weeks only, supports the scientific goal to investigate multiple and different places at similar boundary (= weather) conditions. However, the possible requirement to enter “difficult” terrain like the edges of dune field or potentially soft old river beds must be carefully analysed and traded against the inherent risks.

\textbf{Answer:}

\hspace{3mm}The context provided does not mention the MarsFAST mission. The text appears to be a technical paper discussing various topics such as statistical sampling, Monte Carlo simulations, and uncertainty quantification in space situational awareness (SSA) and space debris mitigation. There is no mention of a specific mission called MarsFAST. If you meant to ask a different question, please feel free to rephrase it, and I'll do my best to assist you.

\end{promptboxcollumn}
\caption[SpaceQA incorrect answer example 4]{Example \#4 of an incorrect answer generated by the model on the SpaceQA dataset. Although the passage clearly describes both the technical and scientific goals of the mission, the model fails to associate “this mission” with MarsFAST and incorrectly concludes that the mission is not mentioned, highlighting a failure in co-reference resolution.}

\label{fig:space_qa_example_4}
\end{figure}


\end{document}